\documentclass[10pt, conference, a4paper]{IEEEtran}

\usepackage{cite}

\ifCLASSINFOpdf
  \usepackage[pdftex]{graphicx}
\else
  \usepackage[dvips]{graphicx}
\fi
\usepackage[tight,footnotesize]{subfigure}
\usepackage{url}

\usepackage[colorinlistoftodos, obeyDraft]{todonotes}

\usepackage{amsmath}

\begin{document}
%
\title{Implementation of the SWIM Mobility Model in OMNeT++}

\author{\IEEEauthorblockN{Asanga Udugama$^{1}$, Behruz Khalilov$^{1}$, Anas bin Muslim$^{1}$, Anna Foerster$^{1}$, Koojana Kuladinithi$^{2}$\vspace{0.2cm}}
\IEEEauthorblockA{$^{1}$Sustainable Communications Networks Group, 
University of Bremen,
Germany\\
Email: \{adu $|$ bk $|$ anas1 $|$ afoerster\}@comnets.uni-bremen.de}

\IEEEauthorblockA{$^{2}$Communications Networks Group, 
Hamburg University of Technology,
Germany\\
Email: \{koojana.kuladinithi\}@tuhh.de}
}


%


\maketitle

\begin{abstract}

The Internet of Things (IoT) is expected to grow into billions of devices in the near future. Evaluating mechanisms such as networking architectures for communications in the IoT require the use of simulators due to the scale of the size of networks. Mobility is one of the key aspects that need to be evaluated when considering the different scenarios of the IoT. Small Worlds in Motion (SWIM) is a mobility model that mathematically characterises the movement patterns of humans. The work presented in this paper details the development and verification of the SWIM mobility model in the OMNeT++ simulator.   
   

\end{abstract}


%
\IEEEpeerreviewmaketitle

\section{Introduction}
\label{sec:intro}

The amount of devices in the Internet of Things (IoT) is expected to be over 200 billion by 2020 \cite{IDC:2012}. Peer-to-peer opportunistic networking is a model that is being considered for communications in the IoT \cite{AF:2015}. When considering  usage scenarios of the IoT based devices, many scenarios relate to human oriented activities (e.g., health monitoring). When the human element is involved, mobility of devices become an inherent characteristic of any scenario (e.g., a Smartphone user who has enabled health monitoring may travel to work daily).

When evaluating the performance of models for communications in the IoT, use of simulators such as OMNeT++ become a necessity due to the scale of the networks. The myriad of devices that form the IoT require appropriate models to correctly evaluate the performance. One such modeling area is mobility.

The OMNeT++ simulation environment provides implementations for a number of mobility models. Though these mobility models were developed to address different requirements of simulating mobile devices, \cite{Mei:2009} has shown through empirical evaluations that mobility patterns of humans are much more different than what these models model. Therefore, to cater to this disparity, the authors of \cite{Mei:2009} have proposed a simple self tuning mathematical model to model human behaviour with respect to mobility.

This model, named Small Worlds In Motion (SWIM) \cite{Mei:2009} uses two intuitions of human mobility, viz., people usually visits mostly locations close to their home location and if they visit a location far from home, it is due to its popularity. The authors of \cite{Mei:2009} have characterised these intuitions in a simple equation that considers the distance to a location and the popularity of that location. The work presented in this paper provides a description and the verification of the implementation of the SWIM mobility model in OMNeT++.

The rest of this paper is ordered in the following manner. The next section (Section~\ref{sec:omnet_mm}) provides a brief introduction to the classifications of the mobility models already supported in OMNeT++ and the implementation architecture of mobility in OMNeT++. Section~\ref{sec:swim_details} provides a description of the SWIM mobility model highlighting the most relevant parts required to implement the model in OMNeT++. Section~\ref{sec:swim_impl} details the implementation of the SWIM model in OMNeT++. Section~\ref{sec:swim_eval} is a validation of the performance of the model in terms of the selections made of destinations. Section~\ref{sec:conclusion} is a concluding summary.

\section{Mobility Models and Mobility Implementation Architecture in OMNeT++}
\label{sec:omnet_mm}

Network simulators employ two methodologies to simulate node mobility, viz., synthetic mobility models and traces \cite{Camp:2002}. Synthetic models are mathematical models that provide a list of coordinates for a node to determine its movement pattern. Traces are coordinate data collected from actual movement patterns of nodes that could be fed into a simulator to simulate mobility.

OMNeT++ provides a number of implementations for synthetic models and trace based mobility. All these implementations extend the functionality of a set of base classes related to handling mobility in OMNeT++. The synthetic mobility models in OMNeT++ are further classified into models with deterministic movements and models with random movements as described below. 


\subsection{Synthetic Mobility Models}

\textbf{Deterministic Movement Models} - Models based on deterministic movements have the characteristic of the same movement pattern without any stochasticity. An example is the \emph{Linear Mobility Model} where a node is made to move at a constant speed within a pre-determined mobility area. When the node reaches a border of the mobility area, it deflects at a pre-determined angle to continue with it's mobility.

%
%
%
%
%

\textbf{Random Movement Models} - Models with random movements employ stochastic movement patterns to move a node within a given area. An example is the \emph{Random Way-point Mobility Model} where a node moves in segments within the mobility area. The speed and the direction for each segment are uniformly distributed random numbers.

%
%
%
%
%
%
%
%
%
%
%
%
%
%
%
%
%
%
%
%
%
%
%
%
%
%
%
%

The SWIM model implemented in this work is a synthetic model with a random movement direction determined by the SWIM equation (described in Section~\ref{sec:swim_details}).

\subsection{Mobility Implementation Architecture in OMNeT++}

OMNeT++ provides an interface (i.e., Abstract Class) that any mobility mechanism (model or trace) must implement to enable mobility in a node. This interface, called \emph{IMobility} provides a set of methods that is invoked from other models that require mobility related information. An example is the wireless propagation models which require to know current coordinates, speed of mobility, etc. to determine the connectivity to other nodes.

To simplify the implementation of mobility, a set of base implementations are provided in OMNeT++ that implements some of the abstract methods of \emph{IMobility}. These base implementations focus on implementing the basic functionality required in mobility patterns that could be generalised into categories. An example is the \emph{LineSegmentsMobilityBase} class which is a general class to be extended when the pattern of mobility is linear and in segments within the mobility area. Similarly, the \emph{LinearRotatingMobilityBase} provides the basic functionality for a node that is required to move in a circular manner within the mobility area.

The SWIM implementation presented in this work extends the \emph{LineSegmentsMobilityBase} class.

\section{Small Worlds in Motion (SWIM)}
\label{sec:swim_details}

Small Worlds In Motion (SWIM) \cite{Mei:2009} is a simple mobility model meant for efficient simulation of human movements. The model is based on the following two intuitions of human movements. 

\begin{itemize}
	\item A visited location is either near to a person's home location;
	\item or, if the visited location is far from the home location, it is visited due to the popularity of that location.
\end{itemize}

The basis for characterising human mobility as itemised above is due to the nature of human mobility as explained below. 

In daily life, people travel more often to places that are near to their homes or to places where they can meet other people. People rarely travel to places which are far from their home locations and if they decide to travel, the reason is due to the popularity of that location. When making travel decisions, it is usually a trade-off; Where to go? Either to go to the best supermarket or the most popular restaurant that is also not far from home location. Very rarely do people travel to a place that is far from home, or that is not popular.

These simple behavioural observations of human mobility provide the basis for SWIM which the authors have validated using real traces and the distributions of key parameters (viz., Inter-contact Time, Contact Duration and the number of contact distributions between nodes).

In the SWIM mobility model, each person (i.e., node) is assigned to a specific home location that is a randomly and uniformly chosen over the network area. Additionally, each node also assigns a weight to each of the possible destinations (i.e., locations). The weight is a function of the popularity of the location and the distance from home location; higher popularities influencing in a positive manner while higher distances influencing in a negative manner. In the SWIM model a network is divided into many small cells that represent possible locations. The size of the cells depend on transmitting range of mobile nodes in that cell. Each node builds a map based on this knowledge. A mobile node calculates the weight for each cell in the map. Once the weight for each cell is known, a node uses this information to choose the next destination to travel to. According to \cite{Mei:2009} the node chooses its destination cell randomly and proportionally considering the $\alpha$ and each node's weight, whereas the exact destination point within the chosen cell is taken uniformly at random over the cell's area. During the movement of a mobile node two characteristics will be followed. Firstly, the speed that a node travels from its current location to destination remains same. Secondly, the movement pattern is a straight line. When a node reaches a destination, it pauses for a period of pause time, chooses the next destination and continues the process. The process of selecting a destination is a two-step process.

\begin{enumerate}
	\item Determine the next destination location type (i.e., a neighbouring location or a remote location). A constant called $\alpha$ is used to determine the destination type (i.e., an $\alpha$ probability of selecting a neighbouring location and a $1 - \alpha$ probability of selecting a remote location);
	\item Once the destination type is selected, the weight is used to determine the exact destination location from the type of destination. The equation for computing the weight is given below. 
\end{enumerate}

\vspace{-0.5cm}

\begin{equation*}
w(C) = \alpha \cdot distance(h_{A}, C) + (1 - \alpha) \cdot seen(C)
\end{equation*}

This equation represents the weight that node $A$ assigns to cell $C$. 

\begin{itemize}
	\item $distance(h_{A}, C)$ is the distance from node $A$ to cell $C$, and it decays based on power-law;
	\item $\alpha$ is a constant in the range of $[0;1]$;
	\item $seen(C)$ is the number of nodes that node $A$ encountered at cell $C$, and this value updates each time node $A$ visits cell $C$.
\end{itemize}
	
The value of $\alpha$ influences the next destination chosen. If the value of $\alpha$ is large, then a mobile node is more likely to choose a destination near to it's home location while a small $\alpha$ results in the node selecting popular locations away from the home location.

The waiting time \cite{Mei:2009} in the SWIM model follows an upper bounded power law distribution (i.e., a truncated power law). This characterisation follows how people usually spend their time; most time at a few locations (e.g., workplace) and short times at many of the other locations (Pareto Principle). 

The distribution of the inter-contact times show power law characteristics until a certain point of time after which it demonstrates an exponential cut-off. Experiments have proven \cite{Mei:2009} that the use of an appropriate $\alpha$ value results in the distribution of inter-contact times as in real life.

\section{SWIM Implementation in OMNeT++}
\label{sec:swim_impl}

The SWIM mobility model implemented for OMNeT++ extends the functionality in the INET framework. The INET framework provides the basic functionality required by SWIM in the following areas.

\begin{itemize}
	\item Constant speed of mobility
	\item Linear movements  	
\end{itemize}

\begin{figure}[!ht]
  \centering
    \includegraphics[width=0.46\textwidth]{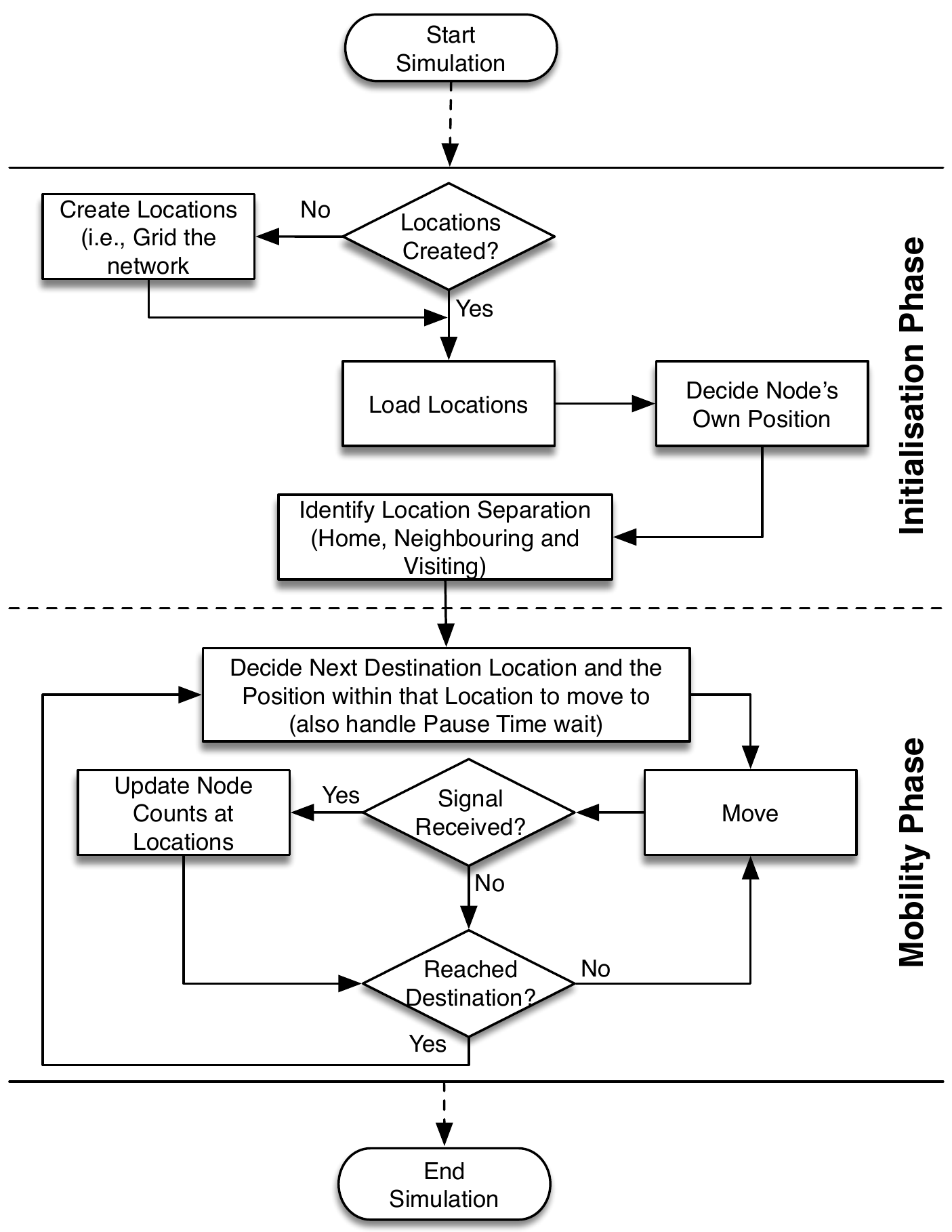}
  \caption{Operation of the SWIM Mobility Model}
  \label{fig:model_operation}
\end{figure}

The \emph{LineSegmentsMobilityBase} of OMNeT++ implements the functionality of linear movements at constant speeds to a given destination. The SWIM extensions implement the decisions on next destination selections. Figure~\ref{fig:model_operation} shows the procedure of the operation of the SWIM mobility model. 

The implementation has two parts to its operation, viz., \emph{Initialisation Phase} and \emph{Mobility Phase}. In the implementation, reference is made to two key terms; \emph{Positions} and \emph{Locations}. A position is the exact coordinates of a node at any given time. A location is an area where multiple nodes may be positioned.
 
\subsection{Initialisation Phase}

The \emph{Initialisation Phase} is implemented in the \emph{initialize()} method of the model implementation architecture of OMNeT++. The following tasks are performed in the initialisation.

\textbf{Create and Load Locations} - The network is identified by a cubic area within which all nodes will move (identified by the coordinates \emph{constraintAreaX}, \emph{constraintAreaY}, \emph{constraintAreaZ}, \emph{constraintAreaWidth}, \emph{constraintAreaHeight} and \emph{constraintAreaDepth}. This area is split into a grid of sub-areas which are identified as \emph{Locations}. The size of each of these locations are equal and they correspond to the wireless range of the wireless technology used to communicate between the nodes. It is assumed, that all nodes in a single location is able to communicate with each other directly. These locations are common to all nodes in the network and therefore, the node that performs the \emph{Initialisation Phase} first, creates a text file of these locations with their coordinates which are then loaded by all the other nodes in their corresponding \emph{Initialisation Phases}.

\textbf{Decide Node's Own Position} - Every node has a starting position identified by the coordinates that lie within the mobility area (identified by \emph{initialX}, \emph{initialY} and \emph{initialZ}). These initial position coordinates are uniformly distributed random numbers (i.e., coordinates) over the mobility area.

\textbf{Identify Location Separation} - Once a node's position is determined and all the locations are identified, the relevance of each location with respect to the node must be determined. The relevance falls into three categories.
\begin{itemize}
	\item \emph{Home} Location - Home location is the area of the grid where the node is located at the beginning of the simulation.
	\item \emph{Neighbouring} Location - A neighbouring location is a location that is in the vicinity of a home location.  Each node will have many neighbouring locations. The determination of neighbouring locations is done by using the \emph{neighbourLocationLimit} model parameter. Any location that lies within the radius (distance given by \emph{neighbourLocationLimit}) of a node's home location is considered as a neighbouring location.  
	\item \emph{Visiting} Location - If a location, does not fall into any of the above two categories, it is considered as a visiting location.
\end{itemize}

\begin{table}[!ht]
	\centering
	\footnotesize
	\begin{tabular}{|p{4cm}|p{4cm}|}
		\hline
		\multicolumn{1}{|c|}{\textbf{Parameter}} & \multicolumn{1}{|c|}{\textbf{Purpose}} \\
		\hline
		\hline
		neighbourLocationLimit & Distance in meters used to determine the location separation (i.e., \emph{Neighbouring} or \emph{Visiting})    \\
		\hline
		 speed                 & Speed of movements in meters per second \\
		\hline
		initialX, initialY, initialZ & Initial coordinates of the node \\
		\hline
		maxAreaX, maxAreaY, maxAreaZ & The area that the node may be allowed to move in, if such a restriction is required to be made \\
		\hline
		waitTime      &  The pause time of the node once it reaches a destination (before it starts moving to the next destination) \\
		\hline
		alpha         & The $\alpha$ value used in the SWIM equation (see Section~\ref{sec:swim_details}) \\
		\hline
		noOfLocations & The number of locations to create from the movement area \\
		\hline
	\end{tabular}
	\vspace{0.1cm}
	\caption{Model Parameters}
	\label{tab:model_parameters}
	\vspace{-0.8cm}
\end{table}

\subsection{Mobility Phase}

The actions performed in the \emph{Mobility Phase} is implemented in the \emph{setTargetPosition()} and the \emph{move()} methods of the mobility related INET classes, and the \emph{receiveSignal} method of the signal mechanism of OMNeT++. The SWIM implementation uses the signal mechanism of OMNeT++ to notify nodes about the arrival of a node at a certain location. The following tasks are performed in the \emph{Mobility Phase}.

\textbf{Decide Next Destination and Pause Time} - When a node reaches a given destination, it has to pause for the specified time which is given with the \emph{waitTime} model parameter. Before pausing, the node sends a signal to other nodes to indicate its arrival at a new location. Once the pause time is completed, the next destination to move to, is identified. The next destination is determined randomly but influenced by $\alpha$ value and the weights assigned to each of the destinations by a node. The equation in Section~\ref{sec:swim_details} computes the weight assigned to each location.

\textbf{Move} - The linear and constant speed movement of the node is handled by the \emph{move()} method of the \emph{LineSegmentsMobilityBase} class which is extended by the implementation.

\textbf{Update Node Counts when Signal Received} - When other nodes send signals indicating their arrival at a certain location, all nodes that are currently in that location must update their node counts to build their location popularity information. Nodes that are not at the location of the originator of the signal will ignore the signal as they have not "\emph{seen}" (i.e., second part of the SWIM equation in Section~\ref{sec:swim_details}) the node.

Table~\ref{tab:model_parameters} shows the configurable parameters of the model which could have static values or based on distributions.

\vspace{-0.15cm}

\section{Model Verification}
\label{sec:swim_eval}

\begin{figure}[!ht]
  \centering
    \includegraphics[width=0.29\textwidth]{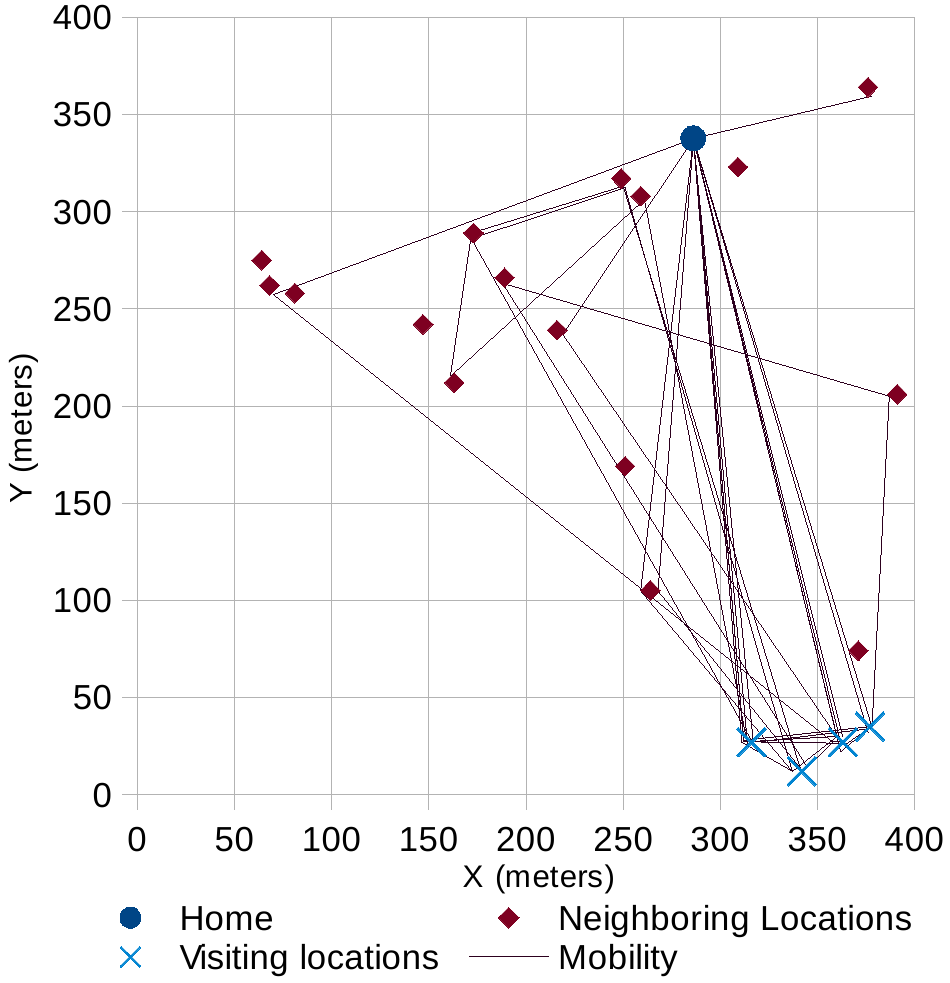}
  \caption{Movement Pattern of a Node with an $\alpha$ of 0.3}
  \label{fig:node_0_alpha_3}
\end{figure}

The performance verification of the movement patterns of the developed SWIM model is done using the example mobility setup in INET. Figure~\ref{fig:node_0_alpha_3} shows the movement pattern of a single mobile node and with an $\alpha$ set to 0.3. The model has identified 21 locations to move between and one of them is the home location. Based on the distance (determined by the \emph{neighbourLocationLimit} model parameter), 16 locations are considered as neighbouring locations while the rest of the locations are considered as visiting locations. As explained before (in Section~\ref{sec:swim_details}), the $\alpha$ value influences the likeliness of the next destination selection. Since the $\alpha$ is a lower value (i.e., 0.3), selection of destinations is skewed towards selecting more remote locations (i.e., visiting locations) than neighbouring locations. 

\begin{figure}[!ht]
  \centering
    \includegraphics[width=0.29\textwidth]{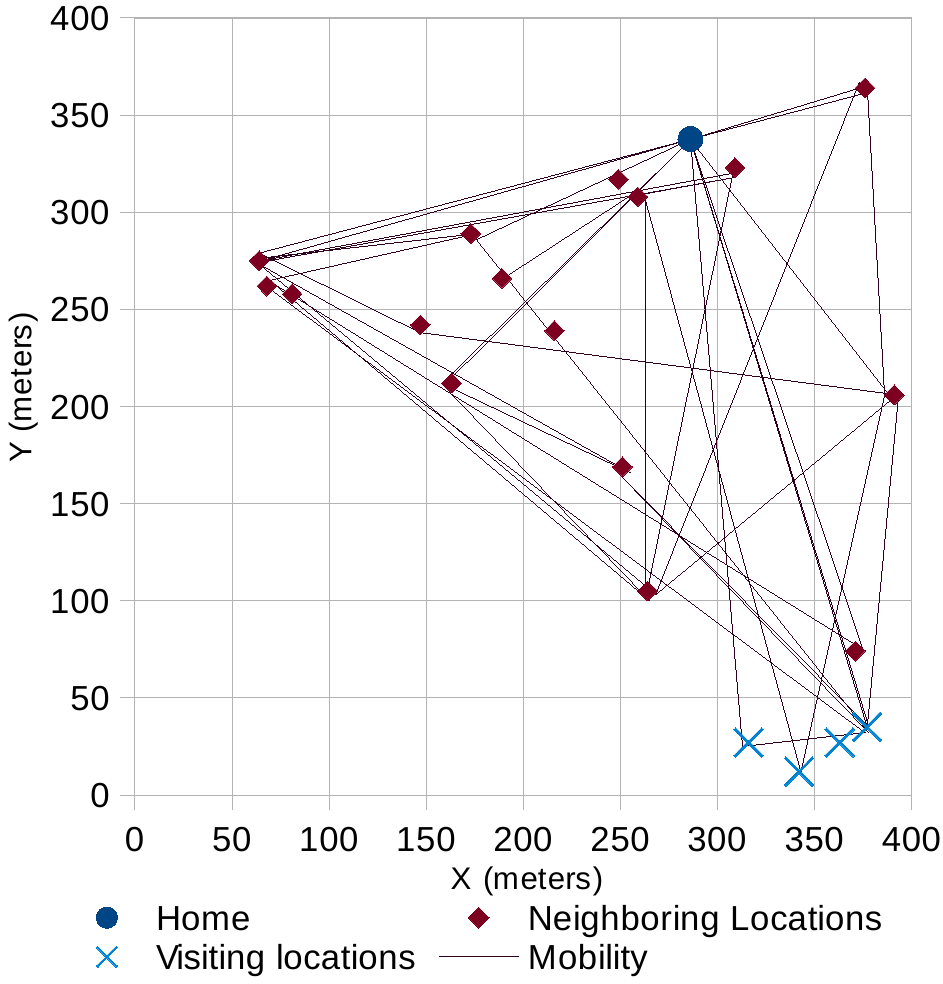}
  \caption{Movement Pattern of a Node with an $\alpha$ of 0.8}
  \label{fig:node_0_alpha_8}
\end{figure}

Figure~\ref{fig:node_0_alpha_8} shows the movement pattern of the same scenario (as in Figure~\ref{fig:node_0_alpha_3}) but with an $\alpha$ of 0.8. Since the $\alpha$ is higher, the location selection is skewed towards selecting more neighbouring locations than remote locations.

Table~\ref{tab:parameter_values} lists the parameters used in obtaining the above shown results. The mobility area was specified using 2 dimensions (i.e., \emph{initialZ} and \emph{maxAreaZ} were set to 0).

\begin{table}[!ht]
	\centering
	\footnotesize
	\begin{tabular}{|p{4cm}|p{4cm}|}
		\hline
		\multicolumn{1}{|c|}{\textbf{Parameter}} & \multicolumn{1}{|c|}{\textbf{Value}} \\
		\hline
		\hline
		neighbourLocationLimit & 300 meters    \\
		\hline
		 speed                 & 1.4 meters per second \\
		\hline
		initialX, initialY     & Randomly selected values \\
		\hline
		maxAreaX, maxAreaY     & 400 meters x 400 meters \\
		\hline
		waitTime               & Randomly selected value between 2 and 5 seconds \\
		\hline
		alpha                  & 0.3 and 0.8 \\
		\hline
		noOfLocations          & 21 locations \\
		\hline
	\end{tabular}
	\vspace{0.1cm}
	\caption{Used Parameter Values}
	\label{tab:parameter_values}
	\vspace{-0.6cm}
\end{table}

\section{Conclusion}
\label{sec:conclusion}

The work presented in this paper relates to the development of the SWIM \cite{Mei:2009} mobility model in OMNeT++ to simulate human behaviour based movements. The work involved in understanding the mathematical model behind the operation of SWIM, understanding the mobility implementation architecture of OMNeT++, implementing the SWIM model in OMNeT++ and finally, verifying the performance.

The verification shows the influence of the differing $\alpha$ values on the location selection process. A larger $\alpha$ value results in a simulation selecting more of the neighbouring locations and less of the remote (i.e., visiting) locations. On the other hand, a smaller $\alpha$ value results in the vise-versa of location selections.

The SWIM mobility model developed in this work is released at Github (\emph{https://github.com/ComNets-Bremen/SWIMMobility}) under a GPL License.

\bibliographystyle{IEEEtran}
%

\end{document}